\documentclass[aps,reprint]{revtex4-1}
\usepackage{amsmath,amsfonts,amssymb}
\usepackage{graphicx,float}
\usepackage{epstopdf}
\usepackage{CJK}

\usepackage[normalem]{ulem}

\usepackage{dcolumn}
\usepackage{bm}
\usepackage{color}

\begin{document}

\preprint{AIP/123-QED}

\title{The Leaky Pipeline in Physics Publishing}

\author{Clara Ross}
 \affiliation{Vassar College.}
\author{Aditya Gupta}%
\author{Ninareh Mehrabi}%
\author{Goran Muric}%
\author{Kristina Lerman}
 \email{lerman@.edu.}
\affiliation{
USC Information Sciences Institute
}%

\date{\today}

\begin{abstract}
Women make up a shrinking portion of physics faculty in senior positions, a phenomenon known as a ``leaky pipeline.'' While fixing this problem has been a priority in academic institutions, efforts have been stymied by the diverse sources of leaks. In this paper we identify a bias potentially contributing to the leaky pipeline. We analyze bibliographic data provided by the American Physical Society (APS), a leading publisher of physics research. By inferring the gender of authors from names, we are able to measure the fraction of women authors over past decades. We show that the more selective, higher impact APS journals have lower fractions of women authors compared to other APS journals. Correcting this bias may help more women publish in prestigious APS journals, and in turn help improve their academic promotion cases.

\end{abstract}

\maketitle

\section{Introduction}
\label{sec-intro}
The gender gap in science have proved remarkably resilient. While girls perform at least as well as boys in secondary school STEM classes~\cite{stoet2018gender}, a gender gap appears in higher education and widens as women decide to pursue careers in science and technology. Women receive 35\% of the bachelor's degrees in STEM~\cite{de2019status}, but only about 20\% of the doctorates. 
Although some progress has been made in recent decades in increasing the representation of women in STEM doctoral programs and entry-level academic positions, such as a lecturer or an assistant professor, representation in senior professorial positions has remained much more stagnant. And despite the progress that has been made, women still represent less than a quarter of the STEM workforce~\cite{hill2010so}.  The problem is especially acute in physics, where among faculty in four-year colleges and universities, women represent 23\% of assistant professors, 18\% of associate professors and just 10\% of full professors~\cite{porter2019women}.

This shrinking participation of women in STEM is known as the "leaky pipeline," and fixing it is essential for the future of scientific innovation. Retaining talented women will increase the diversity of the scientific workforce, a factor associated with higher creativity and productivity in research~\cite{page, smith2017diversity, Woolley2010}. Increasing representation of women, especially in senior academic positions, will also create a self-perpetuating cycle by inspiring more young women to pursue careers in science, and promote gender equality that makes it easier for even more women to ascend academic ranks.

Researchers have identified some of the sources of leaks in the STEM pipeline. The self-confidence gap leads girls to rate their abilities in science and math lower than boys, despite performing similarly well on these subjects in high school~\cite{hill2010so}. As a result, fewer women choose to major in science in college. In fact, low-achieving boys who perform in the 1st percentile on high school math and science assessment tests are just as likely to choose a science major in college as high-achieving girls who score in the 80th percentile~\cite{cimpian2020understanding}.
After they enter the academic workforce, women continue to face more challenges than men, including balancing family responsibilities, an unwelcoming work environment, reduced funding opportunities, and explicit bias~\cite{goulden2009staying,wenneraas2000chair,way2016gender}. The self-confidence gap seen in early education is not just internal; women must meet higher expectations to receive the same opportunities as their male colleagues. One study found that female applicants for an academic position had to be significantly more productive than male applicants to receive a similar rating~\cite{wenneraas1997sexism}. These harms accumulate. Studies have determined that women are more likely than their male colleagues to leave academia earlier and change career paths, despite having equal research productivity and commitment to their careers~\cite{huang2020historical,xu2008gender}. Thus, the loss of women from academia and differences in their accomplishments cannot be justified by their gender or familial responsibilities alone. Female researchers face a multitude of barriers and biases that inhibit their ability pursue the same opportunities as their male colleagues.


In this paper we identify another source of gender disparity in physics, namely a gender gap in physics publishing. We analyze data provided by the American Physical Society (APS), a leading publisher of physics research. APS publishes specialized disciplinary journals, such as Physical Review A (atomic physics) and E (statistical physics), as well as journals aimed at a broader audience, such as its prestigious Physical Review Letters (PRL) and Reviews of Modern Physics. 
We infer the genders of authors based on their names using state-of-the-art methods. We show that the share of women among all authors with identified genders lags far behind that of men. Moreover, the share of women drops with the journal impact factor, and it is lowest for the most prestigious PRL and RMP. Not only do these journals have lower acceptance rates---PRL published only a third of all submitted papers---critically, the editors play a decisive role in which papers are initially reviewed. In contrast, the proportion of female authors is highest in the Physical Review Physics Education Research journal. 
Our findings demonstrate a ``leaky pipeline'' in physics publishing, where the proportion of women authors gradually shrinks with each additional stage of review. Although women publish a representative share of physics research in the less selective APS journals, they are less likely to publish in high-impact APS journals, potentially with negative impact to their careers.
The additional decisions made by editors and reviewers of high impact journals create additional opportunities for implicit bias to creep in. By quantifying sources of bias, we hope to provide the tools for mitigating them to increase representation of women in physics.


\section{Data and Methods}
\label{sec-methods}
The \textbf{APS data} is provided by the American Physical Society (APS), a leading publisher of physics research. The journals published by the APS include the prestigious Physical Review Letters (PRL) and Reviews of Modern Physics (RMP), as well as specialized journals on topics ranging from atomic physics (PRA) and nuclear physics (PRC), to statistical physics (PRE), fluid mechanics (Fluids) and physics education research (PRPER). The APS dataset contains information about more than $450,000$ articles published by the APS journals between the years 1893 and 2018. The data is available at  https://journals.aps.org/datasets. The article metadata include full names of authors, their affiliations, and the year and journal in which the article was published.



To infer the author's gender from the name we use a state-of-the-art service \textit{genderizer.io}~\cite{santamaria2018comparison}. Given a name, it returns a gender and a confidence score between 0.5 and 1. The uncertainty is greater for Asian names that often are not gender-specific. We filter out authors with confidence score less than 0.8, about 19\% of the names, composed predominantly of Chinese and Korean names. In addition, some authors in the APS dataset are identified by their initials only which makes it impossible to detect the gender. Leaving these types of authors out, we performed some general pre-processing techniques on the remaining names, such as normalizing, in order to get the most accurate gender labels for the names in this dataset.

\section{Results}
\label{sec-results}


The number of authors publishing in APS journals has grown greatly over the past decades, 
but despite the growth, women are still a small fraction of all APS authors with identified genders. 
Over the last 50 years, the share of female authors has increased from  4.4\%  in 1970 to 14.8\% in 2018.  These numbers closely mirror the proportion of women physics faculty.

\begin{figure}
    \centering
    \includegraphics[width=0.9\linewidth]{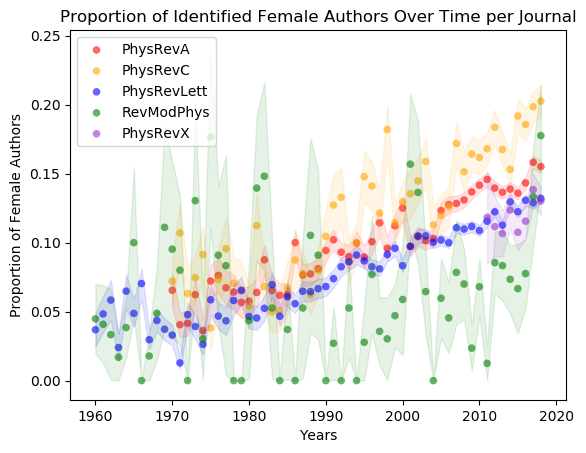}
    \caption{Proportion of female authors over time for selected journals shows systematically lower representation among the authors publishing in Physical Review Letters, Physical Review X and Reviews of Modern Physics.
    }
    \label{fig:journals-years}
\end{figure}

When data is disaggregated by journal, the trends remain remarkably consistent over time. Figure~\ref{fig:journals-years} shows that the proportion of female authors for selected journals grows over time, but those for PRL, PRX and RMP are systematically lower than for other journals.

\begin{figure}
    \centering
        \includegraphics[width=0.9\linewidth]{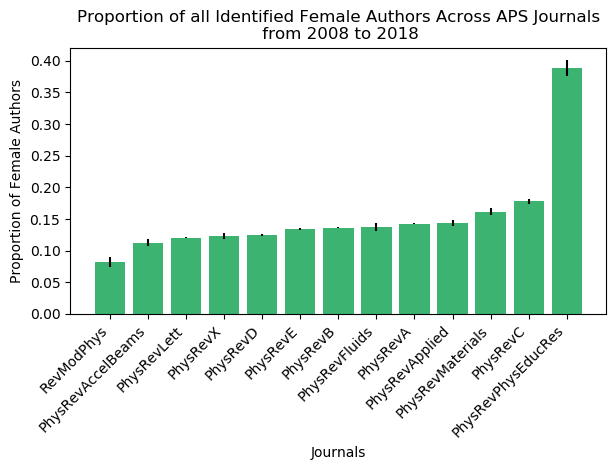}
    \caption{Proportion of female authors publishing in APS journals during the time period 2008--2018. }
    \label{fig:journals}
\end{figure}

\begin{figure}
    \centering
        \includegraphics[width=0.9\linewidth]{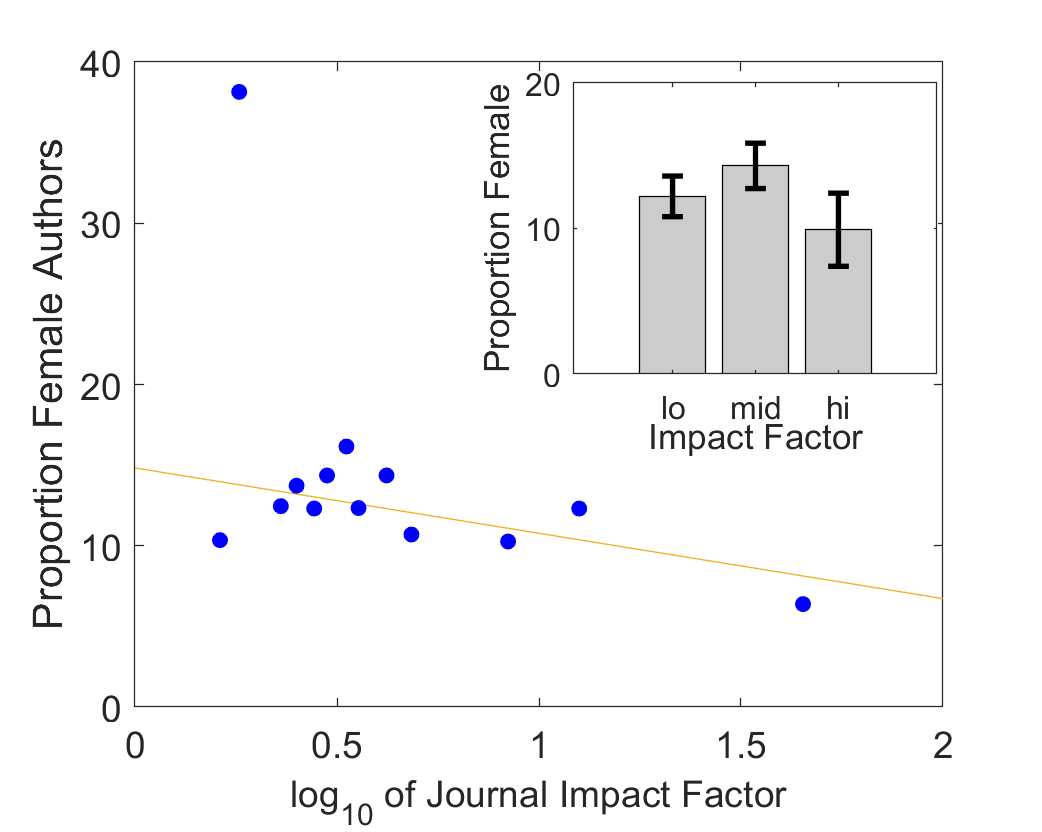}
        \\
        \includegraphics[width=0.9\linewidth]{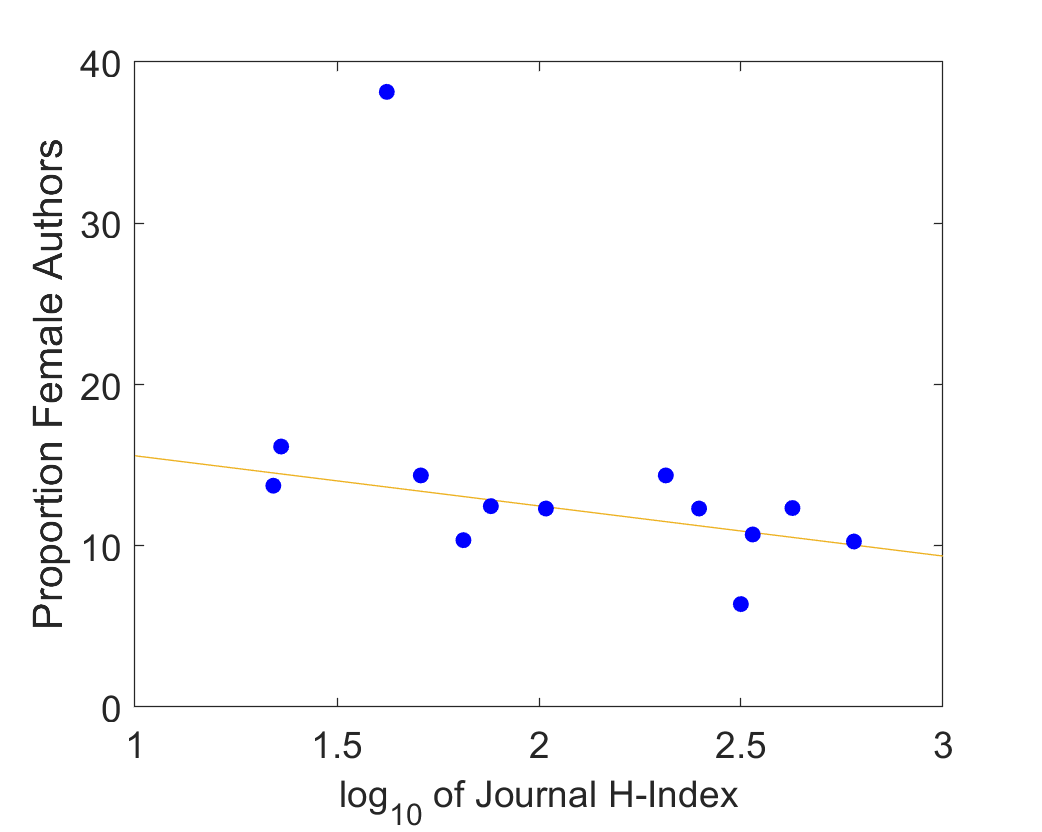}
        \caption{Proportion of female authors as a function of journal impact factor (left) and journal h-index (right) for papers published between 2008 and 2018. The outlier with the high fraction of women is physics education journal. The line shows linear fit and the inset  the average proportion of female authors of journals binned by their impact factor, without the outlier. The bars show standard errors.}
    \label{fig:impact-factor}
\end{figure}

There is large variation in the proportion of female authors when disaggregating data by journal, even after restricting the timescale to the latest decade (Fig.~\ref{fig:journals}). The proportion of female authors is lowest for Review of Modern Physics (8.1\%), Physical Review Accelerators and Beams, Physical Review Letters and Physical Review X. At the other end, the proportion of women authors is highest for Physical Review Physics Education Research journal (38.9\%). 

More prestigious journals publish fewer women authors. Figure~\ref{fig:impact-factor} shows the fraction of women authors publishing in a journal during the decade 2008--2018 as a function of its impact factor and h-index, two common measures of journal prestige. There is a negative correlation between share of women authors and both journal impact factor (Spearman correlation $r=-0.462$, $p=0.11$) and its h-index (Spearman $r=-0.687$, $p=0.01$). Review of Modern Physics has the highest impact factor, i.e., average number of citations per paper, followed by PRX and PRL, while Physics Education Research has one of the lowest (impact factor = 1.811). The differences remain after journals are combined into three groups (after omitting the education outlier): the highest impact tertile that includes the most prestigious journals has significantly  fewer ($p=0.02$) women authors (inset in Fig.~\ref{fig:impact-factor}).
Journal importance is slightly different when judged by h-index (Fig.~\ref{fig:impact-factor}(right)), which measures the number of papers published in the journal with at least that many citations. Here PRL is the most important, followed by PRB, PRD, and then RMP.

\section{Discussion}
Women have made quick gains against the historic discrimination they have faced in higher education.
Although they were not admitted to elite universities until the late 1960's,  by 1980 they surpassed men in the number of bachelors degrees awarded. However, gender equity among STEM degree recipients, as well as among faculty, has remained an elusive goal.
In this study, we have identified a potential factor for this gender gap by demonstrating that a leaky pipeline exists in physics publishing, in which the proportion of female authors publishing physics papers decreases for more prestigious journals. 

We propose that this gender gap may arise from differences in editorial procedures, as our data mirrors the official editorial guidelines for APS journals.
In our study, we found that the Review of Modern Physics has the lowest proportion of female authors. Accordingly, it is the only journal that invites contributions and considers few unsolicited manuscripts. For the next two most impactful journals, PRL and PRX, editors play a major role in deciding whether the paper is sent to referees, who are asked to ``comment critically on the validity and importance of the manuscript,'' according to the APS editorial guidelines. For all the other journals, editors ``select expert(s) in the field to comment on manuscripts that are sent out to review,'' and are not bound to the minimum two referees. The guidelines for these journals are thus more relaxed and involve fewer people reviewing the paper. Our data shows that the proportion of female authors in these journals is higher than that of the first three, and more accurately reflects the fraction of women physicists. 

We suggest two potential reasons for this effect. First, reviewers are subject to bias, which has been responsible for gender gaps across many platforms. Methods to counteract gender biases have historically involved switching to blind review. In one example, the shift to blind auditions among orchestras resulted in a 30\% increase in new female hires~\cite{goldin2000orchestrating}. Each additional stage of the review process adds yet another opportunity for implicit bias to creep in and affect the final decision. Therefore, because more prestigious journals involve more individual reviewers, a leaky pipeline is created.

Another reason for the gender disparities among APS journals may be the norms for rebuttal, and the unwillingness of women to fight for their own work. 
Men may be more likely to appeal a rejection and ultimately see more papers published. In contrast, women are hesitant to engage in  self-promotion~\cite{moss2010disruptions} and as a result, are less likely to appeal an adverse editorial decision. An examination of appeals is needed to shed more light on this question.

The leaky publishing pipeline hurts women's careers and publishers need to take steps to identify and ameliorate gender-based leaks in the research publication pipeline. When women are more likely to publish in prestigious journals, they will be more likely to receive promotions and less likely to drop out of academia. Higher numbers of women faculty members will inspire more young women to choose careers in science, increasing diversity and improving innovation.

%




\bibliographystyle{plain}
\bibliography{references}

\end{document}